\documentclass[aps,prl,twocolumn,superscriptaddress]{revtex4}

\usepackage{graphicx}
\usepackage{dcolumn}
\usepackage{amsmath}
\usepackage{enumerate,amsthm,amssymb,color}
\usepackage{bbm}
\usepackage{bm}
\usepackage[english]{babel}

\begin{document}

\preprint{APS/123-QED}

\title{Experimental demonstration of \\ long-distance continuous-variable quantum key distribution}

\author{Paul Jouguet}\affiliation{LTCI, CNRS - Telecom ParisTech, 46 rue Barrault, 75013 Paris, France}\affiliation{SeQureNet, 23 avenue d'Italie, 75013 Paris, France}
\author{S\'ebastien Kunz-Jacques}\affiliation{SeQureNet, 23 avenue d'Italie, 75013 Paris, France}
\author{Anthony Leverrier}\affiliation{Institute for Theoretical Physics, ETH Zurich, 8093 Zurich, Switzerland}
\author{Philippe Grangier}\affiliation{Laboratoire Charles Fabry de l'Institut d'Optique - CNRS - Univ. Paris-Sud 11, 2 avenue Augustin Fresnel, Campus Polytechnique, 91127 Palaiseau, France}
\author{Eleni Diamanti}\affiliation{LTCI, CNRS - Telecom ParisTech, 46 rue Barrault, 75013 Paris, France}
\maketitle

{\bf Distributing secret keys with information-theoretic security is arguably one of the most important achievements of the field of quantum information processing and communications \cite{SBC:rmp09}. The rapid progress in this field has enabled quantum key distribution (QKD) in real-world conditions \cite{PPA:njp09, SFI:oe11} and commercial devices are now readily available. QKD systems based on continuous variables \cite{WPG:rmp12} present the major advantage that they only require standard telecommunication technology, and in particular, that they do not use photon counters. However, these systems were considered up till now unsuitable for long-distance communication \cite{FDD:njp09,DZV:oe09,JKD:oe12}. Here, we overcome all previous limitations and demonstrate for the first time continuous-variable quantum key distribution over 80 km of optical fibre. The demonstration includes all aspects of a practical scenario, with real-time generation of secret keys, stable operation in a regular environment, and use of finite-size data blocks for secret information computation and key distillation. Our results correspond to an implementation guaranteeing the strongest level of security for QKD reported to date for such long distances and pave the way to practical applications of secure quantum communications.}

Long-distance experiments in quantum information science, and in particular for quantum key distribution (QKD), are of utmost importance for future technological applications. Such experiments will allow the integration of quantum devices in current secure infrastructures and in future networks based on quantum repeaters \cite{BDC:prl98}. The quest for long-distance QKD in the last years has led to several successful demonstrations \cite{TNZ:natphoton07,RPH:njp09,SWV:njp09,DYD:apl10,WCG:ol12}, however improving security guarantees and performance in practical conditions in these implementations remains an issue. These experiments use discrete-variable or distributed-phase-reference protocols \cite{SBC:rmp09}, where the key information is encoded on properties of single photons. Alternatively, in the so-called continuous-variable (CV) QKD protocols \cite{WPG:rmp12}, light carries continuous information such as the value of the quadrature of a coherent state. Such protocols have been implemented in a great variety of situations \cite{GVW:nat03,LRH+:pra06,QHQ+:pra07,LBG:pra07,SAA+:pra07,FDD:njp09,DZV:oe09,SZT+:pra10,JKD:oe12}. Their key feature is that dedicated photon-counting technology can be replaced by homodyne detection techniques that are widely used in classical optical communications. Furthermore, recent results suggest that these systems will be compatible with standard wavelength division multiplexed telecommunication networks \cite{QZQ:njp10}. Despite these attractive features, CVQKD schemes require complex post-processing procedures, mostly related to error correction. These have hindered their operation over more than 25 km of optical fibre \cite{FDD:njp09,DZV:oe09,JKD:oe12}, a communication span that may be insufficient for network cryptographic applications. Furthermore, implementations so far were based on security proofs valid in the asymptotic limit of infinitely large data blocks, while the finite size of real data must be taken into account according to more complete security proofs \cite{SR:prl08,LGG:pra10,JKDL:pra12}.

\begin{figure*}
\centering
 \includegraphics[width=160mm]{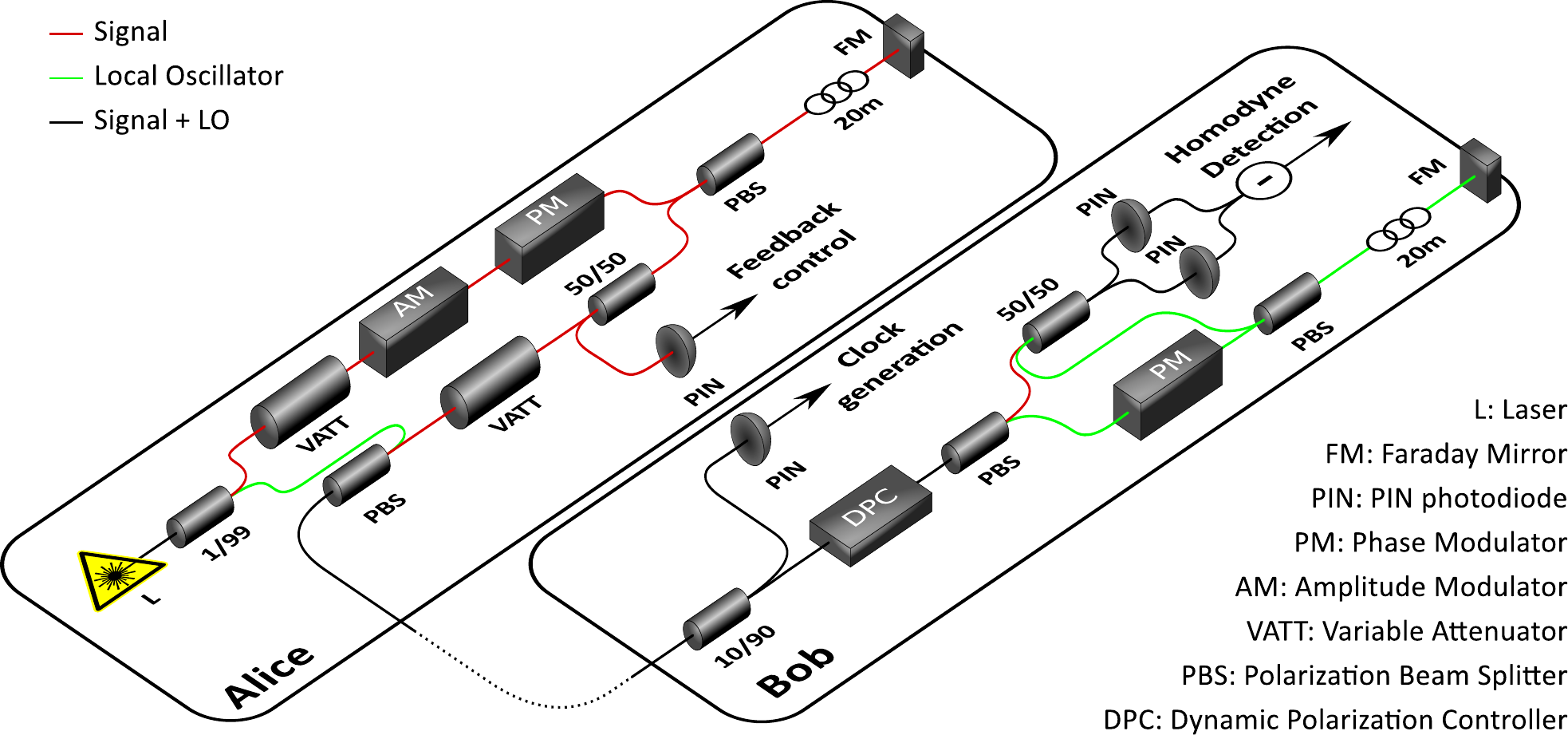}
  \caption{Optical layout of the long-distance CVQKD prototype. Alice sends to Bob 100 ns coherent light pulses generated by a 1550 nm telecom laser diode pulsed with a frequency of 1 MHz. These pulses are split into a weak signal and a strong local oscillator (LO) with an unbalanced coupler. The signal pulse is modulated with a centered Gaussian distribution using an amplitude and a phase modulator. The variance is controlled using a coarse variable attenuator and the amplitude modulator. The signal pulse is 200 ns delayed with respect to the LO pulse using a 20 m delay line and a Faraday mirror. Both pulses are multiplexed with orthogonal polarization using a polarizing beamsplitter (PBS). The time and polarization multiplexed pulses are then sent through the channel. They are demultiplexed on Bob's side with another PBS combined with active polarization control. A second delay line on Bob's side allows for time superposition of signal and LO pulses. After demultiplexing, the signal and LO interfere on a shot-noise limited balanced pulsed homodyne detector. A phase modulator on the LO path allows for random choice of the measured signal quadrature.}
   \label{figure:hardware}
\end{figure*}

Here we demonstrate the distribution of secret keys over a distance of 80 km, using continuous variables and security proofs compatible with the use of data blocks of finite size for the generation of the secret key. This remarkable range improvement was made possible by a system operation in a regime of signal-to-noise ratios (SNR) between one and two orders of magnitude lower than in earlier implementations. In fact, it was previously overlooked that techniques developed for error correction of Gaussian signals \cite{LAB:pra08} can perform very close to the optimal bounds for very low SNR. Implementing such error-correction codes at high speed and in an optical environment featuring excellent stability hence enabling the acquisition of large data blocks, allowed us to reach parameter regions that were previously inaccessible. This was a key element for the present experiments.

In the experimental setup, shown in Figure \ref{figure:hardware}, we implement the standard GG02 coherent-state continuous-variable QKD protocol described in \cite{GG:prl02}. The sender, Alice, prepares coherent states with a Gaussian modulation and sends them to the receiver, Bob, who measures either one of the quadratures with a homodyne detection system. A reverse reconciliation scheme, in which Alice and Bob use Bob's data to establish the secret key \cite{GVW:nat03}, is used. In practice, coherent light pulses with a duration of 100 ns and a repetition rate of 1 MHz are generated by a 1550 nm laser diode and split into a weak signal and a strong local oscillator (LO), which provides the required phase reference. The signal pulses are then randomly modulated in both quadratures using an amplitude and a phase modulator, and transmitted together with the LO pulses in a single fibre using time and polarization multiplexing techniques. At the receiver's side, the pulses interfere on a homodyne detector, whose output is proportional to the signal quadrature selected by a phase modulator placed at Bob's local oscillator path. Due to a highly improved design with respect to previous implementations, the setup presents an excellent stability, ensured by several feedback controls. Polarization drifts occurring in the quantum channel are compensated using a dynamic polarization controller. The beamsplitter placed at the entrance of Bob's apparatus aims at generating from the LO pulse a clock signal that is independent of the polarization state. Then, the homodyne detection statistics and an appropriate algorithm allow us to maintain an optimal polarization state at the output of the channel. The photodiode on Alice's signal path is used for amplitude modulator feedback to correct alterations of the required voltage settings induced by temperature variations, while the homodyne detection output provides the feedback for the phase modulators.

The security of the implemented protocol is well established against collective attacks, both in the asymptotic \cite{GC:prl06,NGA:prl06,PBL:prl09} and in the finite-size regime \cite{LGG:pra10,JKDL:pra12}. Moreover, collective attacks are shown to be asymptotically optimal, thanks, for instance, to an infinite-dimensional version of de Finetti's theorem \cite{RC:prl09}, or to an adaptation of the so-called postselection technique \cite{CKR:prl09} to continuous-variable protocols \cite{LGR:arxiv12}. Security proofs combining arbitrary attacks and finite-size effects are presently actively studied \cite{LGR:arxiv12,FFB:prl12}. Here, we consider the security proofs pertaining to collective attacks \cite{LGG:pra10}, taking also into account finite-size effects. The Gaussian modulation used in the implemented protocol maximizes the mutual information between Alice and Bob, thus offering an optimal theoretical key rate against such attacks. However, it is hard to reconcile correlated Gaussian variables, especially at low signal-to-noise ratios, which are inherent in experiments over long distances. Indeed, the secure distance of previous demonstrations of fiber-based CVQKD \cite{FDD:njp09,JKD:oe12} was limited to about 25 km because no efficient error-correction procedure was available at low SNR. Here, we use the multidimensional reconciliation protocol of \cite{LAB:pra08}, which transforms a Gaussian channel into a virtual binary modulation channel, with a capacity loss that is very low at low SNR. This enables the use of error-correction codes designed for the Binary Input Additive White Gaussian Noise Channel (BIAWGNC) whose typical efficiencies for arbitrarily low SNR are of 0.95 extracted bit per theoretically available bit \cite{JKL:pra11}. This leads to a significant extension of the secure distance.

Let us now look at the parameters that are relevant for the extraction of the secret key. Due to the Gaussian optimality theorem, Alice and Bob's two-mode state at the output of the quantum channel is fully characterized by Alice's modulation variance $V_A$, the channel transmission $T$ and the excess noise $\xi$, which is added by the channel. Both $V_A$ and $\xi$ are expressed in shot noise units. These parameters, together with the shot noise, are estimated in real time using a parameter estimation process, during which a fraction of the samples is randomly chosen and revealed. The other parameters used to compute an estimate of the secret information that can be extracted from the shared data, namely the electronic noise $v_{el}$ and the efficiency of the homodyne detection $\eta$, are measured during a calibration procedure that takes place before the deployment of the system and that is assumed to be performed in a secure environment. Since we have access to high-efficiency error-correction codes only for specific values of the signal-to-noise ratio, the modulation variance $V_A$ is adjusted in real time in order to be at all times as close as possible to the SNR corresponding to the threshold of an available code. When there are several codes available with $V_A$ in its allowed range (roughly between $1$ and $10$ shot noise units on Alice's side), the one corresponding to the higher key rate is chosen.

The privacy amplification step allows us to extract the secret information from the identical strings shared by Alice and Bob after the error-correction procedure. In addition to the amount of data revealed during the error-correction step, we compute an upper bound on the eavesdropper's information on the corrected string for collective attacks in both the asymptotic regime, where all the experimental parameters are assumed to be known with an infinite precision, and in the finite-size regime, where the transmission and the excess noise are estimated over large data pulse sets ($\geq 10^8$ pulses). The stability of our system allows us to obtain a positive secret key rate at long distances in both regimes.

\begin{figure}
\centering
 \includegraphics[width=85mm]{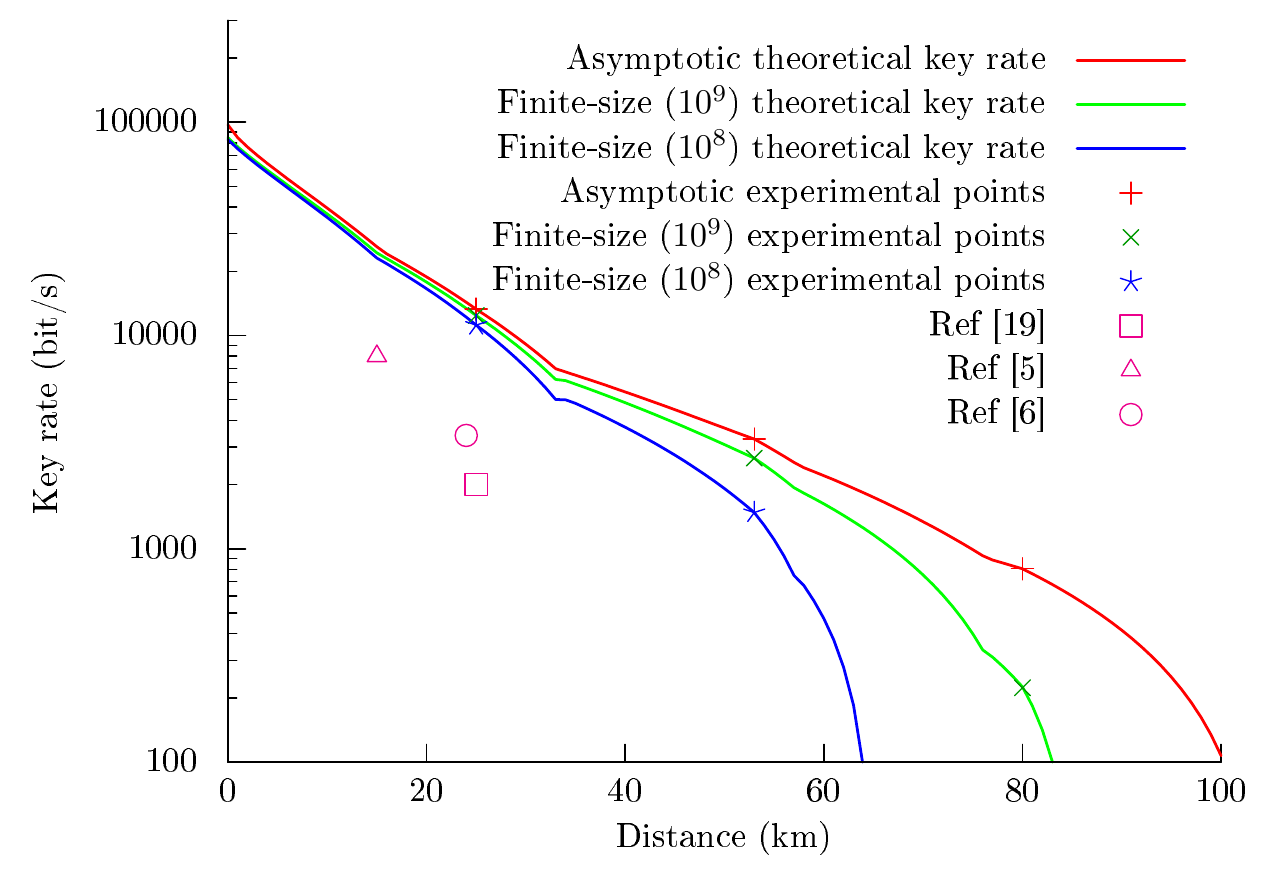}
  \caption{Key rate produced by the system after error correction and privacy amplification during 24 hours with a SNR of 1.1 on Bob's side at 25 km (5.0 dB losses), a SNR of 0.17 at 53 km (10.6 dB losses), and a SNR of 0.08 on Bob's side at 80.5 km (16.1 dB losses). In red, the rate is calculated assuming an eavesdropper able to perform collective attacks in the asymptotic regime. This rate is also valid against arbitrary attacks. In green (resp. blue), the rate is calculated assuming an eavesdropper able to perform collective attacks taking into account finite-size effects with block size $10^9$ (resp. $10^8$) and security parameter $\epsilon = 10^{-10}$. The odd shape of the curves results from the use of a small set of error-correcting codes optimized to perform data reconciliation in specific ranges of SNR. The homodyne detection is characterized by an efficiency $\eta=0.552$, known with uncertainty $\Delta\eta=0.025$, and an electronic noise variance $v_{el}=0.015$, known with uncertainty $\Delta v_{el}=0.002$. For comparison, previous state-of-the-art experimental results are shown \cite{LBG:pra07,FDD:njp09,DZV:oe09}: they are all restricted to distances below 25 km, and do not take finite-size effects into account.}
   \label{figure:key_rate}
\end{figure}

Long-distance secret key generation results for the asymptotic and finite-size regimes are shown in Figure \ref{figure:key_rate}. Secret keys were produced by the experimental system at 25 km (5.0 dB losses), 53 km (10.6 dB losses), and 80.5 km (16.1 dB losses) of standard optical fibre. The key rate was computed during 24 hours at all distances. A sifting procedure reveals $50\%$ of the raw key for parameter estimation, while $50\%$ of the optical pulses have also been discarded for shot noise estimation that is performed in real time. The fraction of light pulses effectively used for generating the key is thus 25 \%. The failure probability of the error correction is roughly 10\%, which corresponds to the amount of sifted key that is discarded. The error correction is performed using Low Density Parity Check (LDPC) codes with a Graphic Processing Unit (GPU) decoder, which features a speed of several Mbits/s \cite{JK:arxiv12}. The final secret key rates range from more than 10 kbits/s at 25 km to a few kbits/s at 53 km and a few hundreds of bits/s at 80.5 km.

\begin{figure}
\centering
 \includegraphics[width=85mm]{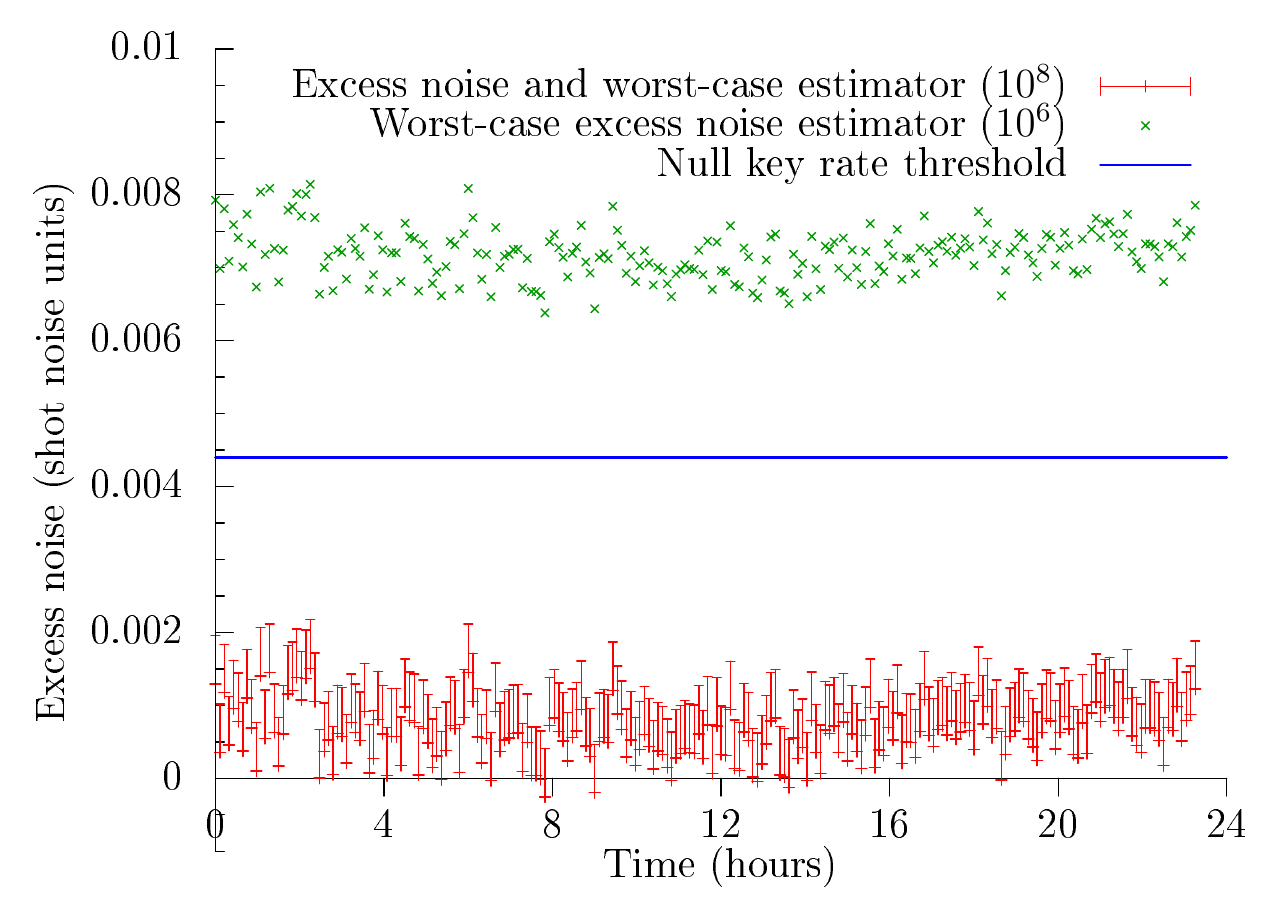}
  \caption{Experimental excess noise measured during 24 hours with a SNR of 0.17 on Bob's side. For this measurement we used 53 km of standard optical fibre corresponding to 10.6 dB losses. The red $+$ correspond to measurements performed on blocks of size $10^8$, each one corresponding to roughly 6 minutes of data acquisition, and indicate the experimental measured excess noise (lower point) as well as the worst-case estimator for the excess noise (upper point) compatible with this data up to a probability of $10^{-10}$. This worst-case estimator is the value used to compute the secret key rate when finite-size effects are taken into account. For comparison, the green $\times$ correspond to the worst-case estimator if the estimation was performed on blocks of size $10^6$. The blue line indicates the maximal value of excess noise that allows for a positive secret key rate. Even without any experimental noise, no secret key could be extracted at 53 km with a parameter estimation on blocks of size $10^6$.}
  \label{figure:excess_noise}
\end{figure}

The results corresponding to the finite-size regime are of particular interest because of their relevance for practical applications. Indeed, obtaining an infinite precision in parameter estimation as required in the asymptotic case is, in practice, impossible. We can further elucidate the results corresponding to the finite-size case by investigating the impact on the secret key rate of the uncertainty on the value of the excess noise. Figure \ref{figure:excess_noise} shows the experimental excess noise measured on blocks of size $10^8$ during 24 hours at a distance of 53 km. Moreover, for each data point, a worst-case estimator of the excess noise compatible with the experimental data is indicated. The probability that the true value of the excess noise is underestimated by the estimator because of statistical fluctuations is less than $10^{-10}$. For comparison, the worst-case estimator for a block size of $10^6$ is also displayed and is clearly incompatible with the extraction of a secret key rate, thus showing that a very large block size is required to achieve long-distance QKD. These results illustrate the significance of the excess noise estimation for system performance. They also confirm the excellent stability of our system, since the excess noise maintains low values, even in this very low SNR regime required by the security proof, and with very large data blocks.

We have demonstrated, for the first time, that long-distance quantum key distribution can be achieved with continuous variables, using only standard telecommunication components. Furthermore, we obtain a positive secret key rate over long distances even when taking into account finite-size effects. This was made possible by exploring a parameter region, namely the region of low signal-to-noise ratios, which was entirely inaccessible in previous implementations. Our results correspond to a practical implementation guaranteeing the strongest level of security reported for QKD at such long distances and show that continuous-variable quantum key distribution can be useful in long-range secure quantum communications and network cryptographic applications.

To conclude, let us discuss possible further improvements of our implementation. The current repetition rate of 1 MHz can be increased by shortening the pulse duration and the time-multiplexing period, as well as the homodyne detection data sampling period, using high speed and high precision data acquisition cards. Current error-correction techniques using GPUs can deal with raw key rates of up to 10 Mbits/s, and even better rates are possible using multiple devices. However, a key rate limitation is imposed by the fact that both the sifting procedure and the multidimensional reconciliation scheme require transmitting a large amount of classical data between Alice and Bob, so increasing the optical rate too much would lead to the saturation of the network link. Finally, the ultimate secure distance that can be reached by our system is determined by the excess noise present in the setup, which sets a fundamental limit to the attainable system performance. In this respect, recent protocols using ``noiseless amplification'' \cite{BLB:pra12} or its ``virtual'' implementation \cite{FC:arxiv12,WSLR:arxiv12} might be promising.

\section{Methods}

{\bf Experimental details:} Our experiment is a one-way implementation, where Alice sends to Bob coherent light pulses with a 100 ns duration and 1 MHz repetition rate generated by a 1550 nm pulsed telecom laser diode. These pulses are split into a weak signal and a strong local oscillator (LO) with an unbalanced coupler. The implemented protocol uses Gaussian modulation of coherent states \cite{GG:prl02}: the signal is randomly modulated in both quadratures using an amplitude and a phase modulator. The signal pulses are then attenuated by a variable attenuator such that the signal power belongs to a range allowing to control the variance of the Gaussian distribution exiting Alice's device using a photodiode and an appropriate feedback algorithm. A second variable attenuator lowers the signal level to a few shot noise units.

The signal and LO are then transmitted through the optical fibre without overlap using time and polarization multiplexing. Delay lines of 200 ns, composed of a 20-m single-mode fibre followed by a Faraday mirror, are used for the time multiplexing. Polarization multiplexing is achieved using polarization beam splitters (PBS). After demultiplexing, the signal and LO interfere on a shot-noise limited balanced pulsed homodyne detector (HD). The electric signal coming from the HD is proportional to the signal quadrature $X_{\phi}$, where $\phi$ is the relative phase between the signal and the LO, which can be controlled using the phase modulator on Bob's LO path according to the Gaussian protocol \cite{GG:prl02}.

Feedback controls are implemented to allow for a stable operation of the system over a large number of pulses ($\geq 10^8$). Polarization drifts occurring in the quantum channel are corrected using a dynamic polarization controller that finds an optimal polarization state at the output of the channel. The photodiode on Alice's signal path is used for the feedback control of Alice's amplitude modulator. On Bob's side, the homodyne detection output is sensitive to phase and can be used to control Alice's and Bob's phase modulators.

{\bf Security conditions:} The implemented protocol requires an exact Gaussian modulation, which is impossible to achieve \cite{JKDL:pra12}. In practice, this is approximated by a truncated discretized modulation with parameters compatible with a security proof against collective attacks. This is done with almost optimal randomness consumption using source coding techniques.

The efficiency and the variance of the electronic noise of the homodyne detection are assumed to be calibrated in a secure laboratory. We evaluate confidence intervals for these values and we compute the eavesdropper's corresponding information taking calibrated values uncertainties into account \cite{JKDL:pra12}.

{\bf Multidimensional reconciliation:} The error-correction step is divided into two parts. First, Bob divides his data into vectors $y$ of size $8$ and for each, draws a binary vector $u$ of the same size at random; $u$ is the reference for the key after the error correction. Then Bob computes $r=y\cdot u$ (where the vectors are interpreted as octonions, see \cite{LAB:pra08} for details) and sends it to Alice who obtains $v=x^{-1}\cdot r = x^{-1}\cdot y \cdot u$, that is a noisy version of Bob's binary modulated vector $u$, with a noise close to a Gaussian noise. Interestingly, it can be shown that the classical data $r$, available to Eve, does not leak any information about the binary vector $u$ \cite{LAB:pra08}. The second step of the error-correction protocol consists in forming vectors of size $2^{20}$ on Alice's and Bob's sides (corresponding to $2^{17}$ pairs of such vectors $u,v$) and to use multi-edge LDPC codes to correct all the errors \cite{JKL:pra11}. The amount of data revealed during this step is subtracted from the secret information previously computed. We use Graphics Processing Unit (GPU) decoding \cite{JK:arxiv12} to obtain a decoding speed compatible with real-time data-processing.

{\bf Post-processing performance:} The error-correction is performed using low-SNR multi-edge Low Density Parity Check (LDPC) codes \cite{JKL:pra11}. High efficiencies are obtained when operating very close to the maximum amount of noise a code can correct. We achieve speeds up to several Mbits/s \cite{JK:arxiv12} using an OpenCL implementation of Belief Propagation with flooding schedule on an AMD Tahiti Graphics Processor. The huge parallelism provided by GPUs allows to overcome the computational complexity of CVQKD pointed out in \cite{ZGC+:ieee08}.

Puncturing and shortening of LDPC codes allow to modify the rate of a code while maintaining a high efficiency over a SNR range. We use a technique proposed in \cite{MEM:qic12} that consists in adapting the code rate with the sum of punctured and shortened bits equal to a constant value determined as the maximum number of punctured bits one can use with a given code without any significant efficiency loss (approximately $10\%$ of the code length). This is particularly convenient since the input size of the error correction is constant; only the output size of the corrected key depends on the number of shortened bits.

Privacy amplification is performed by multiplying by random Toeplitz matrices aggregated corrected key blocks. This can be done very efficiently with input blocks of size $10^9$: a throughput above 40 Mb/s is obtained with one core of a Core i7-920 processor for a rate of $10^{-3}$ secret key bit per raw key bit.

{\bf Hardware stability:} Hardware stability over a long period of time is necessary to extract secret keys from blocks of size greater than $10^9$, as is required to take into account finite-size effects for distances above $50$ km \cite{JKDL:pra12}. A limitation to the long-term hardware stability in \cite{FDD:njp09,JKD:oe12} was the clock reliability on the receiver's side, because of its dependence on the polarization control. We use a splitter and a dedicated electronic circuit and obtain a considerably less noisy and hence more reliable clock.

\section{Acknowledgments}

This research was supported by the French National Research Agency, through the FREQUENCY (ANR-09-BLAN-0410) and HIPERCOM (2011-CHRI-006) projects, and by the European Union through the project Q-CERT (FP7-PEOPLE-2009-IAPP). P.J. acknowledges support from the ANRT (Agence Nationale de la Recherche et de la Technologie). A.L. was supported by the SNF through the National Centre of Competence in Research ``Quantum Science and Technology''.

%


\bibliography{CVQKD_NatPhoton}



\end{document}